\documentclass[journal,twoside,web]{ieeecolor}
\usepackage{generic}
\usepackage{cite}
\usepackage{amsmath,amssymb,amsfonts}
\usepackage{algorithmic}
\usepackage{graphicx}
\usepackage{textcomp}

\def\BibTeX{{\rm B\kern-.05em{\sc i\kern-.025em b}\kern-.08em
    T\kern-.1667em\lower.7ex\hbox{E}\kern-.125emX}}
\begin{document}
\newtheorem{lemma}{Lemma}
\newtheorem{theorem}{Theorem}
\newtheorem{definition}{Definition}
\newtheorem{assumption}{Assumption}
\newtheorem{remark}{Remark}
\newtheorem{algorithm}{Algorithm}
\newtheorem{conjecture}{Conjecture}

\title{\LARGE \bf
Toward a Scalable Upper Bound for a CVaR-LQ Problem
}

\author{Margaret P. Chapman$^\dag$ and Laurent Lessard$^\ddag$ 
\thanks{$^\dag$M.P.C. is with the University of Toronto, Toronto, Canada.\newline Contact email: {\tt\small mchapman@ece.utoronto.ca}}
\thanks{$^\ddag$L.L. is with Northeastern University, Boston, Massachusetts, USA.\newline Contact email: {\tt\small l.lessard@northeastern.edu}}
}

\maketitle
\pagestyle{empty}
\thispagestyle{empty}
\begin{abstract}
We study a linear-quadratic, optimal control problem on a discrete, finite time horizon with distributional ambiguity, in which the cost is assessed via Conditional Value-at-Risk (CVaR). We take steps toward deriving a scalable dynamic programming approach to upper-bound the optimal value function for this problem. \textcolor{black}{This dynamic program yields a novel, tunable risk-averse control policy, which we compare to existing state-of-the-art methods.}
\end{abstract}
\begin{IEEEkeywords}
Stochastic optimal control, LMIs, Linear systems
\end{IEEEkeywords}

\section{Introduction}
\IEEEPARstart{T}{he} standard approach to stochastic optimal control is to evaluate a random cumulative cost in expectation. 
However, this approach is not designed to protect against worst-case circumstances. This limitation motivates robust optimal control \cite{bacsar2008h, zhang2020stability} and related methods, such as minimax model predictive control \cite{robustmpc} and mixed $\mathcal{H}_2/\mathcal{H}_\infty$ control \cite{zhang2020policy}. 

Robust methods typically assume bounded disturbances, which excludes certain common noise models, such as Gaussian noise. A technique to alleviate this restriction is to use a \emph{risk-averse} formulation, in which a random cost is assessed via \emph{exponential utility}. Here, the objective takes the form
    $\mathcal{J}_\gamma(x,\pi) := \textstyle \frac{1}{\gamma} \log\big(E_x^\pi(e^{\gamma Z/2})\big)$,
where $Z \geq 0$ is a random cumulative cost, $\pi$ is a control policy, $x$ is an initial condition, and $\gamma > 0$ is a risk-aversion parameter.\footnote{One may consider $\gamma < 0$, which corresponds to a \emph{risk-seeking} perspective. We focus on the \emph{risk-averse} perspective here, which assumes that noise leads to harm rather than benefit.} This problem has been studied in increasing levels of generality from the 1970s to the 2010s, e.g., see \cite{howardmat1972, jacobson1973, whittle1981risk, whittle1990risk, masi1999, bauerle2014more}. As $\gamma$ increases, the criterion $\mathcal{J}_\gamma(x,\pi)$ represents a more risk-averse perspective, while as $\gamma$ approaches zero, $\mathcal{J}_\gamma(x,\pi)$ tends to the usual expected cost.

In the case of linear dynamics with Gaussian noise and quadratic costs, the problem of optimizing $\mathcal{J}_\gamma(x,\pi)$ is commonly called LEQR control. For a fixed $\gamma > 0$, a Riccati recursion is used to derive the optimal value functions and the optimal control law, which is linear state-feedback \cite{whittle1981risk}. At each step $t$ of the recursion, it must be the case that the matrix $\Sigma^{-1} - \gamma \bar{P}_{t+1}$ is positive definite, where $\Sigma$ is the covariance of the process noise, and $\bar{P}_{t+1}$ is the matrix obtained from step $t+1$. If $\gamma$ is chosen too large, then the above condition may be violated, and the controller synthesis procedure breaks down. While it is known that $\mathcal{J}_\gamma(x,\pi)$ approximates a weighted sum of the expectation $E_x^\pi(Z)$ and the variance $\text{var}_x^\pi(Z)$ if $\gamma \text{var}_x^\pi(Z)$ is ``small'' \cite{whittle1981risk}, a more precise interpretation of $\mathcal{J}_\gamma(x,\pi)$ has not been established.

The \emph{Conditional Value-at-Risk} (CVaR) functional, which was invented in the early 2000s by the financial engineering community \cite{rockafellar2002conditional}, has potential to alleviate the above issues. The CVaR of $Z$ at level $\alpha\in (0,1]$ represents the expectation of the $\alpha \cdot 100\%$ largest values of $Z$.
The intuitive interpretation of CVaR and its quantitative characterization of risk aversion (in terms of a \emph{fraction} of worst-case outcomes) are two reasons for its popularity in financial engineering (see \cite{Kisialadissertation} and the references therein) and its emerging popularity in control (e.g., see \cite{samuelson2018safety, chapmanACC}). In addition to financial applications, CVaR may be a useful tool for the design of stormwater systems \cite{chapmanACC}, which are required to satisfy precise regulatory specifications, and for the operation of robotic systems \cite{majumdar2020should}.

However, the optimization of CVaR is computationally expensive in general. Unlike the expectation of a random (cumulative) cost, the CVaR of a random cost, subject to the dynamics of a Markov decision process, does \emph{not} satisfy a dynamic programming (DP) recursion on the state space. One way to resolve this issue and make DP valid is via a suitable state augmentation~\cite{bauerle2011markov}.

Here, we study a linear-quadratic optimal control problem with \emph{distributional ambiguity}, where the cost is assessed via CVaR. Our first step is to derive an upper bound to the optimal value of this problem.
This derivation (Theorem \ref{mythm1}) and additional analysis (Theorem \ref{thm11}) motivate the formulation of an interesting dynamic programming algorithm (Theorem \ref{thm3}). While the associated value functions are defined on an augmented state space, they are computed in a \emph{scalable} fashion since their parameters come from a Riccati-like recursion. Moreover, our algorithm provides a risk-averse controller, in which a risk-aversion level is parameterized in a novel way through a positive definite matrix. While our controller synthesis procedure is more computationally complex than LEQR, it does not involve a condition that is analogous to the positive definiteness of $\Sigma^{-1} - \gamma \bar{P}_{t+1}$ for all $t$. 
\section{A CVaR-Linear-Quadratic Problem}\label{secII}
\subsection{Notation}
If $M \in \mathbb{R}^{n \times n}$, then $M \geq 0$ ($M > 0$) means that $M$ is symmetric and positive semi-definite (positive definite). Upper-case letters denote random objects (e.g., $X_t$), and lower-case letters denote values of random objects (e.g., $x_t$). If $\mathcal{E}$ is a separable metrizable space, $\mathcal{B}(\mathcal{E})$ is the Borel sigma algebra on $\mathcal{E}$, and $\mathcal{P}(\mathcal{E})$ is the space of probability measures on $(\mathcal{E},\mathcal{B}(\mathcal{E}))$ with the weak topology. 
We define $\bar{\mathbb{R}} := \mathbb{R} \cup \{-\infty,+\infty\}$, $\mathbb{R}_+ := [0,+\infty)$, and $\mathbb{R}_+^{n} := \{ z \in \mathbb{R}^{n} : z_i \in \mathbb{R}_+, \; i = 1,\dots,n\}$. $0_{n \times m}$ is the $n \times m$ zero matrix. $I_{n}$ is the $n \times n$ identity matrix. The trace of a matrix $M \in \mathbb{R}^{n \times n}$ is $\text{tr}(M)$. 
\subsection{Linear-Quadratic System Model}\label{lqsysmodel}
Consider a fully observable, linear time-invariant system:
\begin{equation}\label{sysdyn}
    X_{t+1} = AX_t + BU_t + W_t \; \; \; \forall t \in \{0,1,\dots,N-1\},
\end{equation}
where $X_t$ is a $\mathbb{R}^n$-valued random state, $U_t$ is a $\mathbb{R}^m$-valued random control, and $W_t$ is a $\mathbb{R}^n$-valued random disturbance at time $t$. The matrices $A \in \mathbb{R}^{n \times n}$ and $B \in \mathbb{R}^{n \times m}$ and the length of the time horizon $N \in \mathbb{N}$ are given. The initial state $X_0$ is fixed at an arbitrary $x \in \mathbb{R}^n$. For convenience, define $f(x_t,u_t,w_t) := Ax_t + Bu_t + w_t$ for all $x_t \in \mathbb{R}^n$, $u_t \in \mathbb{R}^m$, and $w_t \in \mathbb{R}^n$.

We make the following assumptions about the $\mathbb{R}^n$-valued disturbance process $(W_0,W_1,\dots,W_{N-1})$. $W_t$ and $W_s$ are independent for all $t \neq s$, and $W_t$ is independent of the initial state $X_0$ for each $t$. For each $t$, the exact distribution of $W_t$ is not known. However, the first and maximal second moment of $W_t$ are known, which we specify below.
\begin{definition}[Ambiguity Set]\label{myPw}
We define $\mathcal{P}_W \subseteq \mathcal{P}(\mathbb{R}^n)$ to be the set of probability measures with zero mean and covariance upper-bounded by $\Sigma > 0$. Each disturbance $W_t$ has a distribution $\nu_t \in \mathcal{P}_W$. In other words, $\nu_t$ satisfies $\textstyle \int_{\mathbb{R}^n} w_t \, \nu_t(\mathrm{d}w_t) = 0_{n \times 1}$ and $\textstyle \int_{\mathbb{R}^n} w_t w_t^T \, \nu_t(\mathrm{d}w_t) \leq \Sigma$.
\end{definition}

As the system evolves, a random cumulative quadratic cost is incurred. The random cost-to-go for time $t \in\{0,1,\dots,N-1\}$ is defined as
\begin{equation}\label{myZ}
    Z_t := \textstyle \underbrace{X_N^T Q_f X_N}_{Z_N} + \sum_{j=t}^{N-1} \underbrace{X_j^T Q X_j + U_j^T R U_j}_{c(X_j,U_j)}.
\end{equation}
$c(X_j,U_j)$ is the random stage cost at time $j$. $Z_N$ is the random terminal cost. $Q \in \mathbb{R}^{n \times n}$, $R \in \mathbb{R}^{m \times m}$, and $Q_f \in \mathbb{R}^{n \times n}$ satisfy $Q > 0$, $R > 0$, and $Q_f > 0$, respectively. We define $Z := Z_0$. With slight abuse of notation, we also use $c(x_t,u_t) = x_t^T Q x_t + u_t^T R u_t$ for all $x_t \in \mathbb{R}^n$ and $u_t \in \mathbb{R}^m$.

\subsection{CVaR-Risk-Averse Optimal Control Problem}
Consider a CVaR optimal control problem on a discrete, finite time horizon with distributional ambiguity:
\begin{equation}\label{cvaropt}
    J_\alpha^*(x) := \inf_{\pi \in \Pi} \; \sup_{\gamma \in \Gamma} \;  \text{CVaR}_{\alpha,x}^{\pi,\gamma}(Z),
 \end{equation}
subject to the linear dynamics \eqref{sysdyn}, where $x \in \mathbb{R}^n$ is an initial condition and $\alpha \in (0,1]$ is a risk-aversion level. The objective $\text{CVaR}_{\alpha,x}^{\pi,\gamma}(Z)$ is the CVaR of $Z$ at level $\alpha$, when the system is initialized at $x$ and evolves according to a control policy $\pi \in \Pi$ and a disturbance strategy $\gamma \in \Gamma$. ($\gamma$ provides a distribution for $W_t$ for each $t$. $\Pi$ and $\Gamma$ will be defined in this section.) The CVaR of $Z$ represents the expectation of the $\alpha \cdot 100$\% largest values of $Z$. 

While the problem \eqref{cvaropt} does not satisfy a dynamic programming (DP) recursion on $\mathbb{R}^n$, there is a useful DP recursion on $\mathbb{R}^n \times \mathbb{R}$ (Lemma \ref{dynprogrammingthm}). A CVaR optimal control problem \emph{without} distributional ambiguity has been solved by defining an augmented state space \cite{bauerle2011markov}. Taking inspiration from \cite{bauerle2011markov}, we use a $\mathbb{R}^n \times \mathbb{R}$-valued, random \emph{augmented state} $(X_t,S_t)$. The dynamics of $X_t$ are given by \eqref{sysdyn}. $S_t$ is a $\mathbb{R}$-valued random variable, whose dynamics are given by
\begin{equation}\label{augsys}
    S_{t+1} = S_t -  c(X_t,U_t) \quad\text{for all } t \in \{0,1,\dots,N-1\}.
\end{equation}
$S_t$ keeps track of the random cumulative cost up to time $t$. The realizations of $(X_0,S_0)$ are concentrated at an arbitrary point $(x,s) \in \mathbb{R}^n \times \mathbb{R}$. We use the \emph{augmented state space} $\mathbb{R}^n \times \mathbb{R}$ to define $\Pi$, the class of history-dependent control policies that summarize the history through $(X_t,S_t)$.
\begin{definition}[Control Policies $\Pi$]\label{controlpolicy}
A control policy $\pi \in \Pi$ takes the form $\pi := (\pi_0,\pi_1,\dots,\pi_{N-1})$,
such that for each $t$, $\pi_t$ is a (Borel-measurable) stochastic kernel on $\mathbb{R}^m$ given $\mathbb{R}^{n}\times \mathbb{R}$.
\end{definition}
%
\begin{definition}[Disturbance Strategies $\Gamma$]\label{disturbancestrategy}
Every disturbance strategy $\gamma \in \Gamma$ takes the form $\gamma := (\nu_0,\nu_1,\dots,\nu_{N-1})$, such that $\nu_t \in \mathcal{P}_W$ is the unknown distribution of $W_t$ for each $t$.
\end{definition}
%
\subsection{Probability Space for Random Cumulative Cost}
For any $(x,s) \in \mathbb{R}^n\times \mathbb{R}$, $\pi \in \Pi$, and $\gamma \in \Gamma$, the random cost $Z = Z_0$ \eqref{myZ} is defined on a probability space $(\Omega,\mathcal{B}(\Omega),P_{x,s}^{\pi,\gamma})$, where
the sample space is $\Omega := (\mathbb{R}^n \times \mathbb{R} \times \mathbb{R}^m)^N \times \mathbb{R}^n \times \mathbb{R}$. Every $\omega \in \Omega$ takes the form
   $ \omega = (x_0,s_0,u_0,\dots,x_{N-1},s_{N-1},u_{N-1},x_N,s_N)$,
where $(x_t,s_t) \in \mathbb{R}^n \times \mathbb{R}$ is the value of $(X_t,S_t)$ and $u_t \in \mathbb{R}^m$ is the value of $U_t$ in the trajectory $\omega$. \textcolor{black}{We have specified implicitly that the coordinates of $\omega$ have causal dependencies via \eqref{sysdyn}, \eqref{augsys}, and Definition \ref{controlpolicy}.} 
The random state at time $t$ is a function $X_t : \Omega \rightarrow \mathbb{R}^n$, such that if $\omega \in \Omega$ is as above, then $X_t(\omega) := x_t$, which is Borel measurable. $S_t$ and $U_t$ are defined analogously. The probability measure $P_{x,s}^{\pi,\gamma}$ is used to evaluate expectations, e.g., $E_{x,s}^{\pi,\gamma}(Z) :=\int_\Omega  Z(\omega) \; \mathrm{d}P_{x,s}^{\pi,\gamma}(\omega)$. 
The form of $P_{x,s}^{\pi,\gamma}$ on measurable rectangles is known, and it depends on the dynamics of the augmented state \eqref{sysdyn} \eqref{augsys}, an initial augmented condition $(x,s) \in \mathbb{R}^n \times \mathbb{R}$, a control policy $\pi \in \Pi$, and a disturbance strategy $\gamma \in \Gamma$ (Ionescu-Tulcea Theorem). For instance, see \cite[Prop. 7.28]{bertsekas2004stochastic} or \cite[Prop. C.10, Remark C.11]{hernandez2012discrete} for details. 
\subsection{Defining CVaR of Random Cumulative Cost}
The Conditional Value-at-Risk of $Z = Z_0$ \eqref{myZ} at a risk-aversion level $\alpha \in (0,1]$ is defined as follows:
\begin{equation}\label{cvardefpi}\begin{aligned}
    \text{CVaR}_{\alpha,x}^{\pi,\gamma}(Z) \!
     := \! \begin{cases}\underset{s \in \mathbb{R}}{\inf} \; g_{\alpha,x}^{\pi,\gamma}(s,Z)  & \hspace{-1mm} \text{if }E_{x,s}^{\pi,\gamma}(Z) < +\infty \; \forall s \in \mathbb{R} \\
    +\infty  & \hspace{-1mm}  \text{otherwise}, \end{cases}
\end{aligned}\end{equation}
where $g_{\alpha,x}^{\pi,\gamma}(s,Z) := s + \textstyle\frac{1}{\alpha} E_{x,s}^{\pi,\gamma}(\max(Z-S_0,0))$.
\begin{remark}
It is standard to define 
\begin{equation*}
    \text{CVaR}_{\alpha}(Y):= \underset{s \in \mathbb{R}}{\inf} \Big( s + \textstyle\frac{1}{\alpha} E(\max(Y-s,0)) \Big),
\end{equation*}
where $Y$ is a random variable such that $E(|Y|) < +\infty$. In \eqref{cvardefpi}, we use an \emph{extended definition} for CVaR to permit a class of policies $\Pi$ that depends on the augmented state space and need not have a particular analytical form (e.g., linear).
\end{remark}

\section{Upper Bound for CVaR-LQ Problem}\label{secIII}
We use the definition of $\text{CVaR}_{\alpha,x}^{\pi,\gamma}(Z)$ \eqref{cvardefpi} to re-express $J^*_\alpha(x)$ \eqref{cvaropt}. For any $x \in \mathbb{R}^n$ and $\alpha \in (0,1]$, it holds that
\begin{equation*}\label{77}\begin{aligned}
  J_\alpha^*(x)
\hspace{-.5mm} & = \hspace{-.5mm} \inf_{\pi \in \Pi}  \sup_{\gamma \in \Gamma} \; \hspace{-1mm}\begin{cases}\underset{s \in \mathbb{R}}{\inf} \; g_{\alpha,x}^{\pi,\gamma}(s,Z) & \hspace{-2mm} \text{if }E_{x,s}^{\pi,\gamma}(Z) \hspace{-.5mm} < \hspace{-.5mm} +\infty \; \forall s \in \mathbb{R} \\
    +\infty & \hspace{-2mm} \text{otherwise} \end{cases}\\
\hspace{-.5mm}  & = \hspace{-.5mm} \inf_{\pi \in \Pi}  \underbrace{\sup_{\substack{\gamma \in \Gamma \\ E_{x,s}^{\pi,\gamma}(Z) < +\infty \; \forall s \in \mathbb{R}}} \; \inf_{s \in \mathbb{R}} \; g_{\alpha,x}^{\pi,\gamma}(s,Z)}_{J_{\alpha,\pi}(x)}.
\end{aligned}\end{equation*}
In the current section, first we show that there is a policy $\pi \in \Pi$ such that $J_{\alpha,\pi}(x)$ is finite (Lemma \ref{lemma111}), which guarantees that the problem \eqref{cvaropt} is well-defined. Second, we derive an upper bound to $J_\alpha^*(x)$ (Theorem \ref{mythm1}):
\begin{equation*}\begin{aligned}
    J_\alpha^*(x) \leq \underset{s \in \mathbb{R}}{\inf} \Big( s + \textstyle \frac{1}{\alpha} V_0^*(x,s) \Big)
\end{aligned}\end{equation*}
with
    $V_0^*(x,s) := \inf_{\pi \in \Pi}  \sup_{\gamma \in \Gamma} E_{x,s}^{\pi,\gamma}(\max(Z-S_0,0))$.
Toward the goal of computing $V_0^*$ scalably, we will define a value iteration algorithm with value functions $V_N,\dots, V_1, V_0$ (Section \ref{secIV}). We will analyze the algorithm in the setting of deterministic policies and finitely many disturbance values. We will show that, under a measurable selection assumption, $\bar{V}_0^*\leq \bar{V}_0$ (Theorem \ref{thm11}), where $\bar{V}_0^*$ and $\bar{V}_0$ are the versions of $V_0^*$ and $V_0$ in the simplified setting, respectively. In Section \ref{secV}, we will prove that $V_0 \leq \hat{V}_0$, where
\begin{equation*}
  \hat{V}_0(x,s) := a_0 + \max(x^T P_0 x - s,0)  \; \; \; \forall (x,s) \in \mathbb{R}^n \times \mathbb{R},
\end{equation*}
such that $a_0 \in \mathbb{R}$ and $P_0 > 0$ are obtained via a Riccati-like recursion (Theorem \ref{thm3}). \textcolor{black}{We will explain how the proof of Theorem \ref{thm3} provides an algorithm for a novel risk-averse controller.} Also, the above analysis takes key steps toward deriving
\begin{equation*}
    J_\alpha^*(x) \leq \underset{s \in \mathbb{R}}{\inf} \Big( s + \textstyle \frac{1}{\alpha}  \hat{V}_0(x,s) \Big)\; \; \; \forall x \in \mathbb{R}^n, \; \forall \alpha \in (0,1],
\end{equation*}
a \emph{scalable} upper bound to a CVaR linear-quadratic optimal control problem with distributional ambiguity.
%
%
\begin{lemma}[$J_{\alpha,\pi}(x)$ is finite for some $\pi$]\label{lemma111}
For all $x \in \mathbb{R}^n$ and $\alpha \in (0,1]$, there is a $\pi \in \Pi$ such that $J_{\alpha,\pi}(x) \in \mathbb{R}$. 
\end{lemma}
\begin{proof}
Let $\pi \in \Pi$ be an open-loop deterministic policy such that $U_t$ takes the value $0_{m \times 1}$ for each $t$. By using the quadratic cost, linear dynamics, and the definition of the ambiguity set, it holds that
\begin{gather}\label{toshowlemma1}
    0 \leq E_{x,s}^{\pi,\gamma}(Z) \leq H_{x}^{\pi,\Sigma} \quad \forall \gamma \in \Gamma, \; \forall s \in \mathbb{R}, 
\end{gather}
where $H_{x}^{\pi,\Sigma} := x^T Q x + \text{tr}\big(Fxx^TF^T\bar{Q}\big) + \text{tr}\big(G\bar{\Sigma} G^T\bar{Q}\big)$.
$\bar{\Sigma}$ is a block diagonal matrix containing $N$ copies of $\Sigma$. $\bar{Q} := \text{diag}(Q, \dots, Q, Q_f)$ is a block diagonal matrix with $N-1$ copies of $Q$. $F \in \mathbb{R}^{Nn \times n}$ and $G \in \mathbb{R}^{Nn \times Nn}$ depend on $A$ and $N$. The desired statement follows from \eqref{toshowlemma1}. 
\end{proof}

By the previous lemma and since $\{ J_{\alpha,\pi}(x) : \pi \in \Pi \}$ is bounded below by 0, it holds that $J_\alpha^*(x) \in \mathbb{R}$.
\begin{theorem}[Upper bound to $J_\alpha^*(x)$]\label{mythm1}
Define
\begin{equation}\label{v0star}\begin{aligned}
    G_{\alpha}(x) & := \inf_{\pi \in \Pi} \; \inf_{s \in \mathbb{R}} \; \sup_{\gamma \in \Gamma} \; g_{\alpha,x}^{\pi,\gamma}(s,Z) \; \; \forall x \in \mathbb{R}^n,  \forall \alpha \in (0,1],\\
    V_0^*(x,s) & \hspace{-1mm} := \hspace{-1mm} \inf_{\pi \in \Pi} \sup_{\gamma \in \Gamma}  E_{x,s}^{\pi,\gamma}(\max(Z\hspace{-1mm} -\hspace{-1mm}S_0,0)) \; \forall x \in \mathbb{R}^n, \forall s \in \mathbb{R}.
\end{aligned}\end{equation}
For all $x \in \mathbb{R}^n$ and $\alpha \in (0,1]$, $J_\alpha^*(x) \leq  G_{\alpha}(x)$, $G_{\alpha}(x) \in \mathbb{R}$, and $G_{\alpha}(x) = \underset{s \in \mathbb{R}}{\inf} ( s + \textstyle \frac{1}{\alpha} V_0^*(x,s) )$. Moreover, $V_0^*$ is finite.
\end{theorem}
\begin{proof}
We have $J_\alpha^*(x) \leq  G_{\alpha}(x)$ because
\begin{equation*}
      \sup_{\gamma \in \Gamma_x^\pi} \; \inf_{s \in \mathbb{R}} \; g_{\alpha,x}^{\pi,\gamma}(s,Z) \leq   \inf_{s \in \mathbb{R}} \; \sup_{\gamma \in \Gamma_x^\pi} \; g_{\alpha,x}^{\pi,\gamma}(s,Z) \; \; \forall \pi \in \Pi, 
\end{equation*}
where $\Gamma_x^\pi := \{ \gamma \in \Gamma : E_{x,s}^{\pi,\gamma}(Z) < +\infty \; \forall s \in \mathbb{R}\}$, and since $\Gamma_x^\pi \subseteq \Gamma$. $G_\alpha(x) \in \mathbb{R}$ because $\{ \sup_{\gamma \in \Gamma} \; g_{\alpha,x}^{\pi,\gamma}(s,Z) : \pi \in \Pi, s \in \mathbb{R}\}$ is bounded below and there
exist $s \in \mathbb{R}$ and $\pi \in \Pi$ such that $\sup_{\gamma \in \Gamma} \; g_{\alpha,x}^{\pi,\gamma}(s,Z) \in \mathbb{R}$. Indeed, let $s=0$, and let $\pi$ assign the value $0_{m \times 1}$ to each $U_t$. Then, we have
\begin{equation*}
    0 \leq g_{\alpha,x}^{\pi,\gamma}(0,Z) = \textstyle\frac{1}{\alpha}E_{x,0}^{\pi,\gamma}(Z) \leq \textstyle\frac{1}{\alpha}H_{x}^{\pi,\Sigma} \; \; \;\; \; \; \forall \gamma \in \Gamma.
\end{equation*}
We have $G_{\alpha}(x) = \inf_{s \in \mathbb{R}} ( s + {\textstyle\frac{1}{\alpha}} V_0^*(x,s))$ because one may exchange the order of infima. $V_0^*$ is finite because 1) for any $(x,s) \in \mathbb{R}^n \times \mathbb{R}$, there is a $\pi \in \Pi$ such that $\sup_{\gamma \in \Gamma}  E_{x,s}^{\pi,\gamma}(\max(Z-S_0,0)) \in \mathbb{R}$, and 2) $\{ \sup_{\gamma \in \Gamma} \; E_{x,s}^{\pi,\gamma}(\max(Z-S_0,0)) : \pi \in \Pi \}$ is bounded below by 0. For the first property, one may choose the policy that assigns the value $0_{m \times 1}$ to each $U_t$. 
\end{proof}

\section{Analysis of a Value Iteration Algorithm}\label{secIV}
To estimate $V_{0}^*$ \eqref{v0star} in a scalable fashion, we propose a value iteration algorithm on $\mathbb{R}^n \times \mathbb{R}$.
\begin{algorithm}[Value Iteration for General Setting]\label{valueiterationalg}
Let the functions $V_N, V_{N-1}, \dots, V_0$ be defined recursively as follows. For all $(x,s) \in \mathbb{R}^n \times \mathbb{R}$ and for $t=N-1,\dots,1,0$,
\begin{align*}
      V_N(x,s) \!&:= \max(x^T Q_f x - s,0)\\
   V_{t}(x,s) \!&:= \!\!\!
   \inf_{u \in \mathbb{R}^m}\!\!
   \sup_{\nu \in \mathcal{P}_W} \textstyle \!\!\!\int_{\mathbb{R}^n}\! V_{t+1}(f(x,u,w), s \!-\! c(x,u)) \nu(\mathrm{d} w).
\end{align*}
\end{algorithm} 
\begin{conjecture}\label{conj1}
\textcolor{black}{The functions $V_{N-1}, \dots, V_1, V_0$ are Borel measurable and bounded below by 0.}
\end{conjecture}

We use the Conjecture in the proof of Theorem \ref{thm3}, which requires the Lebesgue integrals in Algorithm \ref{valueiterationalg} to exist. The Conjecture will be proved formally in future work by using properties of convex functions. 

In this work, we will analyze Algorithm \ref{valueiterationalg} in the setting of finitely many disturbance values and deterministic policies.
\begin{definition}[$\bar{\Pi}$]
$\bar{\Pi}$ is the set of deterministic policies such that every $\pi \in \bar{\Pi}$ takes the form $\pi = (\pi_0,\pi_1,\dots,\pi_{N-1})$, where each $\pi_t : \mathbb{R}^n \times \mathbb{R} \rightarrow \mathbb{R}^m$ is Borel measurable.
\end{definition}
\begin{definition}[$\bar{\mathcal{P}}_W$]
Let $W_t$ be supported on the $N_W \in \mathbb{N}$ points $\{w^{1},w^{2},\dots,w^{N_W}\} \subseteq \mathbb{R}^n$, and let $p_t^{j} \in [0,1]$ be the (unknown) probability that the value of $W_t$ is $w^{j}$. In this case, the ambiguity set of distributions is
\begin{equation*}
    \bar{\mathcal{P}}_W \hspace{-1mm}:= \hspace{-1mm} \left\{ p \in \mathbb{R}_+^{N_W} \;\middle|\; \begin{aligned} & \textstyle \sum_{j=1}^{N_W} p^j = 1, \; \textstyle\sum_{j=1}^{N_W} w^{j} p^{j} = 0_{n \times 1}, \\ & \textstyle \sum_{j=1}^{N_W} w^{j}(w^{j})^T p^{j} \leq \Sigma  \end{aligned}\right\}\hspace{-1mm}.
\end{equation*}
\end{definition}
\begin{definition}[$\bar{\Gamma}$]\label{bargamma}
The set of disturbance strategies in the setting of finitely many disturbance values is
   $ \bar{\Gamma} := \big\{\gamma = (p_0,p_1,\dots,p_{N-1}) : p_t \in \bar{\mathcal{P}}_W \; \forall t \big\}$.
\end{definition}

The version of $V_0^*$ \eqref{v0star} in the setting of finitely many disturbance values and deterministic policies is
\begin{equation}\label{V0bar}
    \bar{V}_{0}^*(x,s) := \inf_{\pi \in \bar{\Pi}} \; \sup_{\gamma \in \bar{\Gamma}} \;  E_{x,s}^{\pi,\gamma}(\max(Z-S_0,0))
\end{equation}
for all $(x,s) \in \mathbb{R}^n \times \mathbb{R}$.
The version of Algorithm \ref{valueiterationalg} in the setting of finitely many disturbance values follows. 
\begin{algorithm}[Value Iteration in Finite Case]\label{valuealg2} Let the functions $\bar{V}_N, \bar{V}_{N-1}, \dots, \bar{V}_0$ be defined recursively as follows. For all $(x,s)  \in \mathbb{R}^n \times \mathbb{R}$ and for $t = N-1,\dots,1,0$,
\begin{align*}
      \bar{V}_N(x,s) \!&:= \max(x^T Q_f x - s,0) \\
    \bar{V}_{t}(x,s) \!&:=\!\!\! \inf_{u \in \mathbb{R}^m}\!\!\! \underbrace{\sup_{p_t \in \bar{\mathcal{P}}_W} \!\!\!\textstyle \sum_{j=1}^{N_W}  p_t^{j} \; \bar{V}_{t+1}(f(x,u,w^{j}), s \!-\! c(x,u))}_{\bar{\psi}_{t+1}(x,s,u)}\!.
\end{align*}
\end{algorithm}

\noindent The next theorem specifies properties of Algorithm \ref{valuealg2}.

\begin{theorem}[Analysis of Algorithm \ref{valuealg2}]\label{thm11}
For $t = 0,1,\dots,N$, the value function $\bar{V}_t : \mathbb{R}^n \times \mathbb{R} \rightarrow \mathbb{R}$ is convex and bounded below by 0, and $\bar{V}_t(x_t,s_t)$ is non-increasing in $s_t$ for each $x_t$. For $t = 0,1,\dots,N-1$, for any $(x_t,s_t) \in \mathbb{R}^n \times \mathbb{R}$, there is a $u^*_{x_t,s_t} \in \mathbb{R}^m$ such that
\begin{equation}
    \bar{V}_{t}(x_t,s_t) = \bar{\psi}_{t+1}(x_t,s_t,u^*_{x_t,s_t}).
\end{equation}
For $t = 0,1,\dots,N-1$, suppose that $\bar{V}_{t}$ is Borel measurable and there is a Borel measurable function $\pi_t^* : \mathbb{R}^n \times \mathbb{R} \rightarrow \mathbb{R}^m$ such that for all $(x_t,s_t) \in \mathbb{R}^n \times \mathbb{R}$,
\begin{equation}\label{vbarpiassumption}
    \bar{V}_{t}(x_t,s_t) = \bar{\psi}_{t+1}(x_t,s_t,\pi_t^*(x_t,s_t)).
\end{equation}
Define $\pi^* := (\pi_0^*,\pi_1^*, \dots, \pi_{N-1}^*)$. Then, Algorithm \ref{valuealg2} provides an upper bound to $\bar{V}_{0}^*$ \eqref{V0bar}, specifically, $\bar{V}_{0}^*\leq \bar{V}_{0}$. 
\end{theorem}
\begin{remark}
Theorem \ref{thm11} invokes a measurable selection assumption (see also \cite[Th. 3.2.1]{hernandez2012discrete}), which motivates future study of measurable selection theorems. 
\end{remark}

To prove Theorem \ref{thm11}, we present two supporting results.
\begin{lemma}[Value Function Analysis]\label{thmanalysis}
Let $v : \mathbb{R}^n \times \mathbb{R} \rightarrow \mathbb{R}$ be convex and bounded below by 0. Also, let $v(x,s)$ be non-increasing in $s$ for each $x$. Define $v^* : \mathbb{R}^n \times \mathbb{R} \rightarrow \bar{\mathbb{R}}$ as
    $v^*(x,s) := \inf_{u \in \mathbb{R}^m}  \sup_{p \in \bar{\mathcal{P}}_W} \textstyle \sum_{j=1}^{N_W} p^{j} \; v\big(f(x,u,w^j), s - c(x,u)\big)$.
Then, $v^*$ is finite, convex, and bounded below by 0, and $v^*(x,s)$ is non-increasing in $s$ for each $x$. Also, for all $(x,s) \in \mathbb{R}^n \times \mathbb{R}$, there is a $u_{x,s}^* \in \mathbb{R}^m$ such that
     $  v^*(x,s) = \sup_{p \in \bar{\mathcal{P}}_W} \textstyle \sum_{j=1}^{N_W} p^{j} \; v\big(f(x,u_{x,s}^*,w^j), s - c(x, u_{x,s}^*)\big)$.
\end{lemma}
\begin{proof}
Since $f(x,u,w^j)$ is affine in $(x,u,s)$ for each $w^j$, $s - c(x,u)$ is concave in $(x,u,s)$, $v$ is convex, and $v(x,s)$ is non-increasing in $s$ for each $x$,
    $(x,u,s) \mapsto v(f(x,u,w^j), s - c(x,u))$
is convex in $(x,u,s)$ for each $w^j$. By proceeding step-by-step through the operations that lead to $v^*$ and by using knowledge of the operations that preserve convexity, the desired properties follow.
\end{proof}

The next supporting result for Theorem \ref{thmanalysis} provides properties of conditional expectations and a DP recursion on $\mathbb{R}^n \times \mathbb{R}$. For $t = 0,1,\dots,N$, the function $\omega \mapsto (X_t(\omega),S_t(\omega))$ is Borel measurable \cite[Prop. 7.14]{bertsekas2004stochastic}. Let $\pi \in \bar{\Pi}$ and $\gamma \in \bar{\Gamma}$ be given. For $t = 0,1,\dots,N$, denote the \emph{$(\pi,\gamma)$-conditional expectation} of $\max(Z_t - S_t,0)$ as follows:
        $\varphi_t^{\pi,\gamma}(x_t,s_t) := E^{\pi,\gamma}(\max(Z_t - S_t,0) | X_t = x_t, S_t = s_t )$,
where $Z_t$ is defined by \eqref{myZ} and $Z = Z_0$. The function $\varphi_t^{\pi,\gamma} : \mathbb{R}^n \times \mathbb{R} \rightarrow \bar{\mathbb{R}}$ is Borel measurable, and $\varphi_t^{\pi,\gamma}$ is almost-everywhere unique with respect to $P_{t,x,s}^{\pi,\gamma} \in \mathcal{P}(\mathbb{R}^{n} \times \mathbb{R})$, which is defined by
\begin{equation}\label{myPtxs}
    P_{t,x,s}^{\pi,\gamma}(K) := P_{x,s}^{\pi,\gamma}\big(\{\omega \in \Omega : (X_t(\omega),S_t(\omega)) \in K \}\big)
\end{equation}
for every $K \in \mathcal{B}(\mathbb{R}^{n} \times \mathbb{R})$
\cite[Th. 6.3.3]{ash1972probability}. 
\begin{lemma}[A DP Recursion]\label{dynprogrammingthm}
Let $\pi  \in \bar{\Pi}$, $\gamma \in \bar{\Gamma}$, and $(x,s) \in \mathbb{R}^n \times \mathbb{R}$ be given. Then, we have
    \begin{equation*}\begin{aligned}
        \varphi_0^{\pi,\gamma}(x,s) & = E^{\pi,\gamma}_{x,s}( \max(\textstyle Z - S_0, 0) ).
    \end{aligned}\end{equation*}
In addition, we have
\begin{equation*}
    \varphi_N^{\pi,\gamma}(x_N,s_N)  =  \max(\textstyle x_N^T Q_f x_N - s_N, 0)
\end{equation*}
for almost every $(x_N,s_N) \in \mathbb{R}^n \times \mathbb{R}$ with respect to $P_{N,x,s}^{\pi,\gamma}$. Lastly, it holds that
\begin{equation*}
\begin{aligned}
    & \varphi_t^{\pi,\gamma}(x_t,s_t) \\ & \textstyle =  \sum_{j=1}^{N_W} p_t^j \;  \varphi_{t+1}^{\pi,\gamma}\big(f(x_t,\pi_t(x_t,s_t),w^j), s_t- c(x_t,\pi_t(x_t,s_t))\big)
\end{aligned}
\end{equation*}
for almost every $(x_t,s_t) \in \mathbb{R}^n \times \mathbb{R}$ with respect to $P_{t,x,s}^{\pi,\gamma}$ and for every $t \in \{N-1,\dots,1,0\}$.
\end{lemma}
\begin{proof} The conclusions follow from the same arguments that are used to prove the DP recursion for expected cumulative costs (when one uses the probability measure $P_{x,s}^{\pi,\gamma}$ and the dynamics of the augmented state).
\end{proof}

Next, we use Lemma \ref{thmanalysis} and Lemma \ref{dynprogrammingthm} to prove Theorem \ref{thm11}.

\begin{proof}[Theorem \ref{thm11}]
The properties of $\bar{V}_t$ hold by verifying the properties of $\bar{V}_N$ and by applying Lemma \ref{thmanalysis} inductively. Next, we show the last statement, i.e., $\bar{V}_0^* \leq \bar{V}_0$. By Lemma \ref{dynprogrammingthm}, we have
        $\bar{V}_{0}^*(x,s) = \inf_{\pi \in \bar{\Pi}}  \sup_{\gamma \in \bar{\Gamma}}   \varphi_0^{\pi,\gamma}(x,s)$ for all $(x,s) \in \mathbb{R}^n \times \mathbb{R}$.
Let $(x,s) \in \mathbb{R}^n \times \mathbb{R}$, $\gamma \in \bar{\Gamma}$, and $t \in \{0,1,\dots,N\}$ be given. It suffices to show that $\varphi_t^{\pi^*,\gamma} \leq \bar{V}_{t}$ almost everywhere with respect to $P_{t,x,s}^{\pi^*,\gamma}$.
Indeed, the above statement implies that $\varphi_0^{\pi^*,\gamma} \leq \bar{V}_{0}$ almost everywhere with respect to $P_{0,x,s}^{\pi^*,\gamma}$. 
It follows that
\begin{equation}\label{my1313}
    \varphi_0^{\pi^*,\gamma}(x,s) \leq \bar{V}_{0}(x,s).
\end{equation}
Since $\gamma \in \bar{\Gamma}$ in \eqref{my1313} is arbitrary, we have 
\begin{equation}
    \sup_{\gamma \in \bar{\Gamma}} \varphi_0^{\pi^*,\gamma}(x,s) \leq \bar{V}_{0}(x,s).
\end{equation}
Then, since $\pi^* \in \bar{\Pi}$ and by the definition of the infimum,
\begin{equation}\label{2622}
  \bar{V}_{0}^*(x,s) :=\inf_{\pi \in \bar{\Pi}}  \sup_{\gamma \in \bar{\Gamma}}  \varphi_0^{\pi,\gamma}(x,s) \leq \sup_{\gamma \in \bar{\Gamma}} \varphi_0^{\pi^*,\gamma}(x,s) \leq \bar{V}_{0}(x,s).
\end{equation}
Since $(x,s) \in \mathbb{R}^n \times \mathbb{R}$ in \eqref{2622} is arbitrary, the proof would be complete. 

We will prove that $\varphi_t^{\pi^*,\gamma} \leq \bar{V}_{t}$ almost everywhere with respect to $P_{t,x,s}^{\pi^*,\gamma}$ by backwards induction on $t$. The base case ($t=N$) holds by Lemma \ref{dynprogrammingthm} and the definition of $\bar{V}_{N}$. Now, assume (the induction hypothesis) that for some $t \in\{ N-1,\dots,1,0\}$ we have
      $ \varphi_{t+1}^{\pi^*,\gamma} \leq \bar{V}_{t+1} $
almost everywhere with respect to $P_{t+1,x,s}^{\pi^*,\gamma}$. For brevity, we use the notation
\begin{equation}
    \chi_{\pi_t^*}^j(x_t,s_t) := \big(f(x_t,\pi_t^*(x_t,s_t),w^j), s_t- c(x_t,\pi_t^*(x_t,s_t))\big).
\end{equation}
By applying Lemma \ref{dynprogrammingthm}, $\bar{V}_{t+1}$ being Borel measurable, and \eqref{vbarpiassumption}, it suffices to show that
\begin{equation}\begin{aligned}
  \textstyle \sum_{j=1}^{N_W} p_t^j \;  \varphi_{t+1}^{\pi^*,\gamma}\big(\chi_{\pi_t^*}^j(x_t,s_t)\big)
 \leq \textstyle \sum_{j=1}^{N_W} p_t^j \;  \bar{V}_{t+1}\big(\chi_{\pi_t^*}^j(x_t,s_t)\big)
\end{aligned}\end{equation}
for almost every $(x_t,s_t) \in \mathbb{R}^n \times \mathbb{R}$ with respect to $P_{t,x,s}^{\pi^*,\gamma}$. This follows from the induction hypothesis, the Borel measurability of $\bar{V}_{t+1}$, and a classic integration result \cite[Th. 1.6.6 (b)]{ash1972probability}. 
This also involves expressing $P_{t+1,x,s}^{\pi^*,\gamma}$ in terms of $P_{t,x,s}^{\pi^*,\gamma}$; the reader may see \cite[p. 192, Eq. (8)]{bertsekas2004stochastic} for a related derivation.
\end{proof}
%
%
\section{A Scalable Upper Bound}\label{secV}
Here, we return to the setting where there may be uncountably many disturbance values. We will derive a scalable upper bound to $V_0$ (Alg. \ref{valueiterationalg}) of the form,
    $\hat{V}_0(x,s) := a_0 + \max(x^T P_0 x - s,0)$ for all $(x,s) \in \mathbb{R}^n \times \mathbb{R}$,
where $a_0 \in \mathbb{R}$ and a positive definite symmetric matrix $P_0 \in \mathbb{R}^{n \times n}$ are obtained through a Riccati-like recursion. The recursion is parameterized by a positive definite symmetric matrix $L$ and provides a risk-averse controller. After the proof of Theorem \ref{thm3}, we will describe the controller synthesis procedure.

\begin{theorem}\label{thm3}
Define $P_N := Q_f$ and $a_N := 0$. Let $L \in \mathbb{R}^{n \times n}$ satisfy $L > 0$. For $t = N-1,\dots,1,0$, define the matrices $P_t \in \mathbb{R}^{n \times n}$, such that $P_t > 0$, and the scalars $a_t \in \mathbb{R}$ recursively,
\begin{align}
    P_t & := A^T\left( P_{t+1}^{-1} + B R^{-1} B^T - (P_{t+1} + L)^{-1}\right)^{-1} A + Q,\notag\\
    a_t & := a_{t+1}+\text{tr}(\Sigma(P_{t+1} + L)).\label{rec}
\end{align}
For all $t \in \{N,\dots,1,0\}$, define
    $\hat{V}_t(x_t,s_t) := a_t + \max(x_t^T P_t x_t - s_t,0)$ for all $(x_t,s_t) \in \mathbb{R}^n \times \mathbb{R}$.
Then, for all $t \in \{ N,\dots,1,0\}$, we have $V_t \leq \hat{V}_t$, 
\textcolor{black}{provided that $V_t$ is Borel measurable and bounded below by 0}.
\end{theorem}
\begin{remark}[About $L$, $P_t$, $a_t$]
$P_t$ and $a_t$ \eqref{rec} are parameterized by $L$. In the finite-time case above, $L \in \mathbb{R}^{n \times n}$ is only required to be symmetric and positive definite.
\end{remark}
%
%
%
\begin{proof}
We proceed by induction. The base case holds because $P_N = Q_f$ and $a_N = 0$. Now assume that for some $t \in \{ N-1,\dots,1,0 \}$, for all $(x_{t+1},s_{t+1}) \in \mathbb{R}^n \times \mathbb{R}$, we have
    $V_{t+1}(x_{t+1},s_{t+1})  \leq a_{t+1} + \max(x_{t+1}^T P_{t+1} x_{t+1} - s_{t+1},0)$,
where $P_{t+1} \in \mathbb{R}^{n \times n}$ satisfies $P_{t+1} >0$ and $a_{t+1}$ is a scalar.
It suffices to show that
$V_{t}(x_{t},s_{t}) \leq a_t + \max(x_t^T P_t x_t - s_t,0) \; \forall (x_{t},s_{t}) \in \mathbb{R}^n \times \mathbb{R},$
where $a_t$ and $P_t$ are defined by \eqref{rec}. Let $(x_{t},s_{t}) \in \mathbb{R}^n \times \mathbb{R}$. Since $\hat{V}_{t+1}$ and \textcolor{black}{$V_{t+1}$ are Borel measurable} and $\textcolor{black}{0 \leq V_{t+1}} \leq \hat{V}_{t+1}$, it holds that
$ V_{t}(x_t,s_t)  
    \leq  a_{t+1} + \inf_{u_t \in \mathbb{R}^m} \sup_{\nu_t \in \mathcal{P}_W} \int_{\mathbb{R}^n}  \max(\phi_{t+1,u_t}^{x_t,s_t}(w_t),0)  \nu_t(\mathrm{d} w_t)$,
where
$\phi_{t+1,u_t}^{x_t,s_t}(w_t)
:= f(x_t,u_t,w_t)^T P_{t+1} f(x_t,u_t,w_t) + c(x_t,u_t) - s_t$.
By weak duality (e.g., see \cite[Lem.~A.1]{zymler2013distributionally}),
\begin{equation}\label{my177}
     V_{t}(x_t,s_t)  
   \leq a_{t+1} + \underbrace{\inf_{u_t \in \mathbb{R}^m} \inf_{M \in \mathcal{M}_{t+1,u_t}^{x_t,s_t}} \text{tr}(\Delta M)}_{\psi(x_t,s_t)},
\end{equation}
where $\Delta := \text{diag}(\Sigma,1)$ and $\mathcal{M}_{t+1,u_t}^{x_t,s_t}$ is the set of matrices $M = \begin{bmatrix} M_{11} & M_{12} \\ M_{12}^T & M_{22} \end{bmatrix} > 0$ s.t. $M_{11} \in \mathbb{R}^{n \times n}$, $M_{22} \in \mathbb{R}$, and
\begin{equation}\label{266}
    \begin{bmatrix} w_t^T & 1\end{bmatrix} M \begin{bmatrix} w_t^T & 1\end{bmatrix}^T > \phi_{t+1,u_t}^{x_t,s_t}(w_t) \; \; \; \forall w_t \in \mathbb{R}^n.
\end{equation}
By matrix algebra, it follows that \eqref{266} is equivalent to
\begin{align}\label{mylmi}
    & \Phi_{x_t,s_t}^M + \bar{Q}^Tu_t\bar{P} + (\bar{Q}^Tu_t\bar{P})^T > 0,\;\;\text{where}\\
    \Phi_{x_t,s_t}^M & := \hspace{-1mm}\addtolength{\arraycolsep}{-1mm}\begin{bmatrix} M - G_{s_t} & {K^{x_t}}^T \\ K^{x_t} & H^{-1}\end{bmatrix}\!,\, 
     K^{x_t} \hspace{-1mm}:= \hspace{-1mm}\begin{bmatrix} I_n & 0_{n \times 1} \\ 0_{(n+m)\times n} & \begin{bmatrix} x_t \\ 0_{m \times 1} \end{bmatrix}\end{bmatrix}\hspace{-1mm}, \notag \\
       G_{s_t}  &  :=  \begin{bmatrix} 0_{n\times n} & 0_{n \times 1} \\ 0_{1 \times n} & -s_t \end{bmatrix}\hspace{-1mm}, \; \bar{Q}^T  := \begin{bmatrix} 0_{(3n+1) \times m} \\ I_m  \end{bmatrix}\hspace{-1mm},\notag\\
       \bar{P} & := \begin{bmatrix} 0_{1 \times n} & 1 & 0_{1 \times (2n+m)}\end{bmatrix}\hspace{-1mm},\notag\\
        H  &  :=  \begin{bmatrix}  P_{t+1} & P_{t+1} \begin{bmatrix}A & B \end{bmatrix} \\  (*)^T  &  \hspace{-1mm} {\textstyle\begin{bmatrix}A  & B \end{bmatrix} ^T \hspace{-1mm} P_{t+1} \hspace{-1mm} \begin{bmatrix}A & B \end{bmatrix} + \text{diag}(Q,R)} \end{bmatrix}.\notag
\end{align}
Here, $(*)$ denotes the appropriate terms for symmetry. 
By \cite[Lemma 3.1]{gahinet1994linear}, \eqref{mylmi} is solvable for $u_t \in \mathbb{R}^m$ if and only if $W_{\bar{P}}^T \Phi_{x_t,s_t}^M W_{\bar{P}} > 0$ and $W_{\bar{Q}}^T \Phi_{x_t,s_t}^M W_{\bar{Q}} > 0$,
where the columns of $W_{\bar{P}}$ and $W_{\bar{Q}}$ form bases for the nullspaces of $\bar{P}$ and $\bar{Q}$, respectively. 
By matrix algebra, it holds that
\begin{gather}\begin{aligned}\label{103}
    W_{\bar{P}}^T \Phi_{x_t,s_t}^M W_{\bar{P}} > 0 & \iff M_{11} > P_{t+1},\\
    W_{\bar{Q}}^T \Phi_{x_t,s_t}^M W_{\bar{Q}} > 0 & \iff M > H_{x_t,s_t},\;\;\text{where}
\end{aligned}\\
    H_{x_t,s_t} := \begin{bmatrix} \tilde{G} & \tilde{G} A x_t \\ x_t^T A^T \tilde{G} & x_t^T  \big( A^T \tilde{G} A + Q \big) x_t - s_t\end{bmatrix},\;\text{and}\notag \\
    \label{mygtilde}
    \tilde{G} := P_{t+1} - P_{t+1}B( R + B^T P_{t+1}B)^{-1}\! B^T P_{t+1}.
\end{gather}
Therefore, $\psi(x_t,s_t)$ \eqref{my177} is equivalent to
\begin{gather}
    \psi(x_t,s_t) =\inf_{M \in \mathcal{M}_{t+1}^{x_t,s_t}}  \text{tr}(\Delta M),\;\;\text{where} \label{resultpsi}\\
    \mathcal{M}_{t+1}^{x_t,s_t} \hspace{-.5mm} := \hspace{-.5mm} \left\{ \begin{aligned} & M = \begin{bmatrix} M_{11} & M_{12} \\ M_{12}^T & M_{22} \end{bmatrix}  \\ & M_{11} \in \mathbb{R}^{n \times n}\\
    & M_{22} \in \mathbb{R} \end{aligned}\;  \middle|  \;\begin{aligned} M & > 0 \\  M_{11} & > P_{t+1} \\ M & > H_{x_t,s_t} \end{aligned} \right\}.
    \label{myMwithoututequiv}
\end{gather}
By \eqref{my177}, \eqref{resultpsi}, and $\Delta = \text{diag}(\Sigma,1)$, it holds that
\begin{equation}\label{114}
    V_{t}(x_t,s_t) \leq a_{t+1} + \underset{ M \in \mathcal{M}_{t+1}^{x_t,s_t}}{\inf} \text{tr}(\Sigma M_{11}) + M_{22}.
\end{equation}
By taking a Schur complement, $M > H_{x_t,s_t}$ is equivalent to $M_{11} > \tilde{G}$ and $M_{22} > h(x_t,s_t,M)$, where
\begin{multline}\label{myhM}
  h(x_t,s_t,M) := x_t^T  \big( A^T \tilde{G} A + Q \big) x_t 
   - s_t \\
   + (*)^T (M_{11} - \tilde{G})^{-1} \big(M_{12}-\tilde{G} A x_t \big).
\end{multline}
We have $\tilde G \leq P_{t+1}$ from~\eqref{mygtilde}, so $M_{11} > \tilde G$ is redundant:
\begin{equation}\label{myMwithoututequiv2}
    \mathcal{M}_{t+1}^{x_t,s_t} \!=\! \left\{ \begin{aligned} & M \!=\!
    \begin{small}\addtolength{\arraycolsep}{-2pt}\begin{bmatrix} M_{11} & M_{12} \\ M_{12}^T & M_{22} \end{bmatrix}\end{small}  \\ & M_{11} \in \mathbb{R}^{n \times n} \\ & M_{22} \in \mathbb{R} \end{aligned}  \middle|\, \begin{aligned} M & > 0 \\  M_{11} &> P_{t+1} \\  M_{22} & > h(x_t,s_t,M)\! \end{aligned} \right\}\!.
\end{equation}
To bound the objective, we use the relaxation $M_{12}=0_{n \times 1}$.
Recall that $L > 0$ and define the set $\hat{\mathcal{M}}_{t+1,L}^{x_t,s_t} \subseteq \mathcal{M}_{t+1}^{x_t,s_t}$ as:
\begin{equation}\label{MLset}
    \hspace{1pt}\hat{\mathcal{M}}_{t+1,L}^{x_t,s_t} \!:=\! \left\{ \begin{aligned} & M \!=\! \begin{small}\addtolength{\arraycolsep}{-4pt}\begin{bmatrix} M_{11} & 0_{n \times 1} \\ 0_{1 \times n} & M_{22} \end{bmatrix}\end{small}  \\ & M_{11} \in \mathbb{R}^{n \times n} \\ & M_{22} \in \mathbb{R} \end{aligned}  \middle|\, \begin{aligned} M & \!>\! 0 \\  M_{11} & \!>\! P_{t+1} + L\\  M_{22} & \!> \! \hat{h}(x_t,s_t,M_{11})\! \end{aligned} \right\}\!,\hspace{-7pt}
\end{equation}
where we define $\hat{h}(x_t,s_t,M_{11}) := $
\begin{equation}\label{myhhat}
x_t^T \big( A^T \big(\tilde{G}^{-1} - M_{11}^{-1}\big)^{-1} A + Q\big) x_t - s_t.
\end{equation}
Thus, we have
\begin{equation}\label{123}
    V_{t}(x_t,s_t)  \leq a_{t+1} + \underbrace{\underset{ M \in \hat{\mathcal{M}}_{t+1,L}^{x_t,s_t}}{\inf} \text{tr}(\Sigma M_{11}) + M_{22}}_{\phi_L(x_t,s_t)}.
\end{equation}
Let $M_{11}^* := P_{t+1}+L$, $M_{22}^* := \max\big(\hat{h}(x_t,s_t,M_{11}^*),0\big)$, and $M_{x_t,s_t}^* := \text{diag}(M_{11}^*,M_{22}^*)$. Then,
\begin{equation}
\phi_L(x_t,s_t) \leq \text{tr}(\Sigma M_{11}^*) + \max\big(\hat{h}(x_t,s_t,M_{11}^*),0\big).\end{equation}
By substituting the definition of $M_{11}^*$, we have
\begin{equation}
V_{t}(x_t,s_t)
\leq  a_t + \max\big(\hat{h}(x_t,s_t,P_{t+1}+L),0\big),
\end{equation}
where $a_t$ is given by \eqref{rec}. Since $\hat{h}(x_t,s_t,P_{t+1}+L) = x_t^T P_t x_t - s_t$, where $P_t$ is given by \eqref{rec}, we are done.
\end{proof}

\begin{remark}[Controller Synthesis]
Based on the proof of Theorem \ref{thm3}, we can derive a sub-optimal policy as follows. For a fixed $L > 0$, compute the matrices $P_t$ via the recursion \eqref{rec}. Let $x_0 \in \mathbb{R}^n$ be an initial condition. Define $s_0 := x_0^T P_0 x_0$, which depends on $L$ through $P_0$. For $t = 0,1,\dots,N-1$, proceed through the following steps:
\begin{enumerate}
    \item Compute $M^*_{x_t,s_t}$ as per the proof of Theorem \ref{thm3}, $M^*_{x_t,s_t} := \text{diag}(M_{11}^*, M_{22}^*)$, where $M_{11}^* := P_{t+1} + L$, $M_{22}^* := \max(\hat{h}(x_t,s_t,M_{11}^*),0)$, and $\hat{h}$ is given by \eqref{myhhat}. 
    \item Choose a $u_t \in \mathbb{R}^m$ that satisfies \eqref{mylmi} when $M  =   M^*_{x_t,s_t}$. Such a $u_t$ is guaranteed to exist from the choice of $M = M^*_{x_t,s_t}$ and by repeating several steps in the proof above.
    We note that the $u_t$ satisfying~\eqref{mylmi} may not be unique.
    \item Nature chooses a disturbance value $w_t \in \mathbb{R}^n$. 
    \item Calculate $x_{t+1} = Ax_t + Bu_t + w_t$ and $s_{t+1} = s_t - c(x_t,u_t)$. Update $t$ by 1. Go to step 1 if $t < N$.
\end{enumerate}
\end{remark}

We now identify some interesting similarities and differences between our approach and classical methods.
\begin{remark}[Relation to LEQR and LQ games]
The Riccati recursion for the LEQR problem in finite time takes the form \cite{whittle1981risk}: for $t = N-1,\dots,1,0$, 
\begin{equation}\label{my29}
    \bar{P}_t = A^T \big( \bar{P}_{t+1}^{-1} +  B R^{-1} B^T - \gamma \Sigma \big)^{-1} A + Q,
\end{equation}
provided that $\gamma > 0$ is chosen so that $\Sigma^{-1} - \gamma \bar{P}_{t+1}$ is positive definite for each $t$. Similarly, the Riccati recursion for a soft-constrained LQ game takes the form \cite[Eq. 3.4a', p. 53]{bacsar2008h}: for $t = N-1,\dots,1,0$, 
\begin{equation}\label{my30}
    \hat{P}_t = A^T \big( \hat{P}_{t+1}^{-1} + BR^{-1}B^T - \textstyle \frac{1}{\lambda^2} \Sigma \big)^{-1} A + Q,
\end{equation}
provided that $\hat{P}_t$ is invertible for each $t$, $R = I_m$, and $\lambda$ is a scalar parameter representing a disturbance-attenuation level.
The key differences between \eqref{rec}, \eqref{my29}, and \eqref{my30} appear in the terms $\gamma \Sigma$, $\frac{1}{\lambda^2} \Sigma$, and $(P_{t+1} + L)^{-1}$, respectively. Our recursion \eqref{rec} encodes a risk-aversion level through the matrix $(P_{t+1} + L)^{-1}$, whereas the classical recursions \eqref{my29} \eqref{my30} encode risk aversion by scaling the covariance $\Sigma$.
\end{remark}
\begin{remark}[Relation to minimax MPC]
One may interpret an LEQR controller in a model-predictive-control (MPC) setting as an approximate solution to minimax MPC \cite[p. 99]{robustmpc}. In minimax MPC, a matrix $\mathcal{T} \geq 0$, which depends on a bounded region containing the process noise, appears in the algorithm that provides an optimal control \cite[Eq. 8.29, p. 99]{robustmpc}. Our recursion \eqref{rec} has a similar structure since it is parameterized by a matrix $L >0$, and it is plausible that a preferable choice of $L$ depends on the maximal covariance $\Sigma$ (a topic for future investigation). A key distinction between minimax MPC and our approach is the uncertainty model of the process noise. Our approach permits process noise with an unbounded support and a spectrum of possibilities that occur with various probabilities. However, minimax MPC permits process noise that lives in a bounded region with known bounds \cite[p. 42]{robustmpc}. The ``better'' uncertainty model may be application-dependent.
\end{remark}
\section{Numerical Simulation}
\begin{figure}[th]
\centerline{\includegraphics[width=\columnwidth]{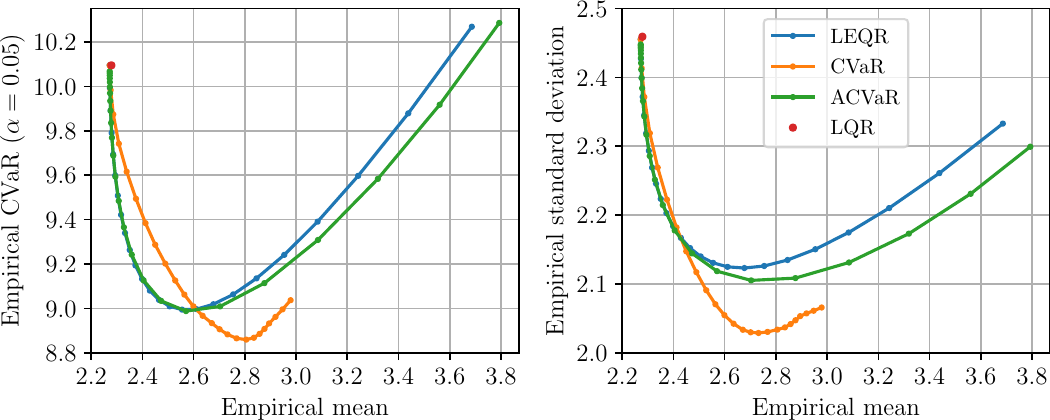}}
\caption{Trade-offs between empirical mean, standard deviation, and $\text{CVaR}_{0.05}$ of the LQR cost for
(i) our controller (\textbf{ACVaR}) as $L$ varies,
(ii) the LEQR controller as $\gamma$ varies (\textbf{LEQR}),
(iii) the exact $\text{CVaR}_{\alpha}$ controller with prior knowledge of the disturbance distribution as $\alpha$ varies (\textbf{CVaR}), and (iv) the LQR controller (\textbf{LQR}).
We used the scalar dynamical system $x_{t+1} = x_t + u_t + w_t$ with $R=Q_f=1$, $Q=10^{-3}$, $x_0=1$, and $N=4$. The disturbance $w_t$ is zero-mean Gaussian with unit variance. The parameter ranges were $0.2 \leq L \leq 100$, $\frac{\gamma_\text{c}}{10} \leq \gamma \leq \gamma_\text{c}$, where $\gamma_\text{c}$ is the critical $\gamma$ value for LEQR. Each point is the mean of 50,000 trials, where the same schedule of pseudo-random seeds are used across policies.
In the limits $L\to\infty$, $\gamma\to 0$, and $\alpha\to 1$ for ACVaR, LEQR, and CVAR$_\alpha$, respectively, we recover the risk-neutral LQR policy.}
\label{exptfigure}
\end{figure}

Fig. \ref{exptfigure} provides example trade-off curves comparing LEQR (as $\gamma$ varies) with our proposed approach from Section~\ref{secV} (as $L$ varies).
These results show that for a simple one-state system, our proposed approach (ACVaR) has comparable performance relative to LEQR.
This finding is notable given the simplicity of our experiment and that our method avoids the case where $\gamma$ is too large and the LEQR cost becomes infinite.
We also simulated the optimal CVaR controller from~\cite{bauerle2011markov}, which is not distributionally robust. This controller assumes exact prior knowledge of the disturbance distribution, which explains its superior performance. However, this optimal CVaR controller is not scalable to higher-dimensional problem instances, since it requires discretizing the augmented state space.

\section{Concluding Remarks}\label{secVI}
%

We took steps toward deriving a scalable upper bound to a distributionally robust, CVaR optimal control problem for linear systems with quadratic costs. CVaR characterizes the (usually abstract) notion of risk as a fraction of worst-case outcomes, which is intuitive and precise. A result from our analysis is a risk-averse controller with intriguing similarities and differences relative to the state-of-the-art.

Potential areas for future work include studying the infinite-horizon case, characterizing the extent to which the upper bound approximation parameterized by $L$ is tight, and elucidating the connections between the choice of $L$ and the maximal covariance $\Sigma$.

Further numerical experiments, potentially with higher-dimensional or more realistic application-specific examples, are needed to ascertain whether the proposed approach may be a superior alternative to LEQR in certain application domains.

%
%
\section*{Acknowledgment}
Both authors would like to thank the reviewers for their valuable feedback regarding connections to existing literature in robust and risk-sensitive control. M.P.~Chapman acknowledges support from the University of Toronto. L.~Lessard is partially supported by NSF awards 1710892 and 1750162.

\bibliographystyle{IEEEtran}
\bibliography{references_new}

\end{document}